\documentclass{elsart}
\usepackage{graphicx}

\newcommand{\be}{\begin{equation}}
\newcommand{\ee}{\end{equation}}
\newcommand{\bea}{\begin{eqnarray}}
\newcommand{\eea}{\end{eqnarray}}

\newcommand{\BE}{\begin{eqnarray}}
\newcommand{\EE}{\end{eqnarray}}
\newcommand{\BEn}{\begin{eqnarray*}}
\newcommand{\EEn}{\end{eqnarray*}}
\newcommand{\barr}{\begin{array}}
\newcommand{\earr}{\end{array}}

\newcommand{\bit}{\begin{itemize}}      
\newcommand{\eit}{\end{itemize}}
\newcommand{\bc}{\begin{center}}
\newcommand{\ec}{\end{center}}
\newcommand{\ben}{\begin{enumerate}}    
\newcommand{\een}{\end{enumerate}}

\newcommand{\bR}{\ensuremath{\mathbf{R}}}
\newcommand{\bomega}{{\mbox{\boldmath $\omega$}}}
\newcommand{\bxi}{{\mbox{\boldmath $\xi$}}}
\begin{document}

\begin{frontmatter}
\title{The Minority Game: a statistical physics perspective}
\author{David Sherrington*}
\address{Rudolf Peierls Centre for Theoretical Physics, University of Oxford, 1 Keble Road, Oxford OX1 3NP, United 
Kingdom 
\\
{\mdseries}* E-mail address: {\tt sherr@thphys.ox.ac.uk}}
 
\date{\today}

%%%%%%%%%%%%%%%%%%%%%%%%%%%%%%%%%%%%%%%%%%%%%%%%%%%%%%%%%%%%%%%%%%%%%%%%%%%%%

\begin{abstract}
A brief review is given of the minority game, an idealized model stimulated by a
 market of speculative agents, and its complex many-body
 behaviour. 
Particular consideration is given to analytic results for the model rather than 
discussions of its relevance in real-world situations.

\end{abstract}
\end{frontmatter}


%\narrowtext 

There is currently much interest in the statistical physics community
 in the emergence of complex co-operative behaviour as a consequence of 
frustration and disorder in systems of simple microscopic constituents
 and simple rules of interaction
\cite{N}. Appropriate minimalist models, designed to capture the essence of real
world problems without peripheral complications have 
played crucial roles in the understanding of such systems. The
 minority game is a such a minimalist model, 
introduced in econophysics to mimic a market of speculators trying to profit by 
buying low and selling high. 
In this paper we review it from the perspective of statistical physics, with a 
view to exposing relevant cooperative and complex features, to demonstrate 
the significant, but incomplete, degree of analytic solubility currently achievable, and
 to illustrate the possibilities for potentially soluble extensions.  

The model describes a system of a large number $N$ of agents each 
of whom at each step of a discrete dynamics makes a bid that can be either positive or negative (buy or 
sell). The objective of each agent is to make a bid of opposite
 sign from that of the sum of all the bids ({\it i.e.} a minority choice). No agent has any direct 
knowledge of the actions or propensities of the others but is aware of
 the cumulative action (total bid) made at each step. Each agent decides his/her bid 
through the application of a personal strategy operator to
 some common information, available identically to all. In the simplest
 versions of the model, to which we restrict here, the strategy 
operators are allocated randomly and independently for each agent before 
play commences and are not modified during play. Each agent has a finite set of 
strategies, one of which is chosen and used at each step. The choice 
is determined by `points' allocated to the strategies and augmented
 regularly via a comparison between the bid associated with 
 playing the stategy and the actual total bid, being increased
 for minority prediction. This is the only mechanism for co-operation 
but is sufficient to yield complex macroscopic behaviour. 

In the original version of the model \cite{CZ} the information on which
 decisions were made was the history of the actual play over a finite 
window (the last $m$ time steps). However, simulations demonstrated that 
utilising instead a random fictitious `history' (information) at each time-step 
produces essentially identical behaviour, suggesting that its
 relevance is just to provide a mechanism for an effective 
 interaction between agents.
 A natural non-trivial measure of the macroscopic behaviour is the
 volatility, the standard deviation of the total bid. It
  demonstrates statistical physics 
 interest in several ways: (i) in exhibiting non-trivial 
 scaling behaviour as a 
 function of $d=D/N$, where $D$ is 
the information dimension \cite{CZ2}, (ii) in exhibiting a 
cusp at a critical $d_c$ following a {\em tabula rasa} start , and 
especially (iii) in that the system is ergodic with volatility 
independent of starting point allocations 
for $d > d_c$ but non-ergodic and preparation-dependent for $d < d_c$; 
see fig 1. This is reminiscent 
of the susceptibility of an infinite-range spin glass 
where a critical temperature $T_c$ separates a preparation-dependent 
regime from an  equilibrating one. 

\begin{figure}
\centerline{\includegraphics[width=33pc]{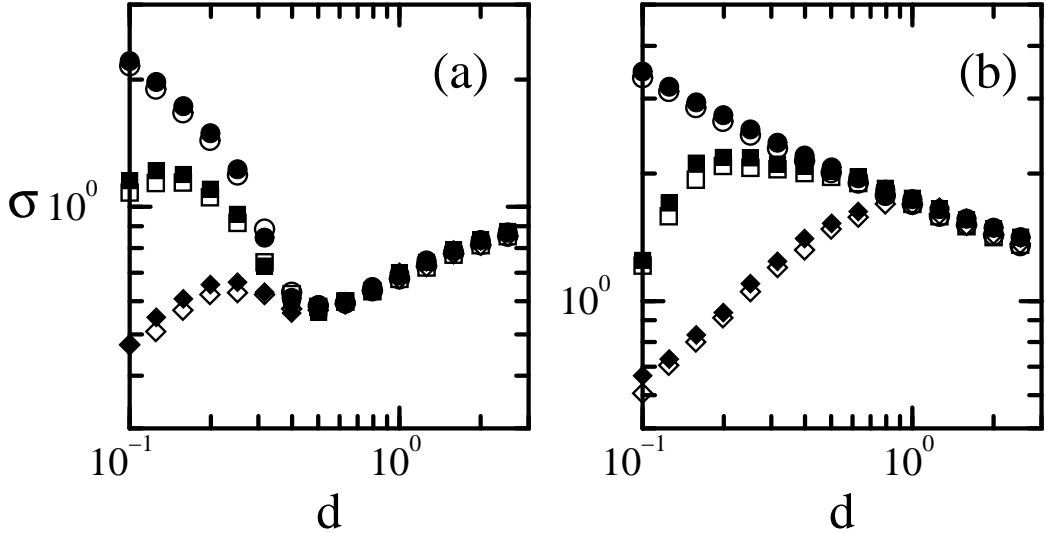}}
\caption{Volatilities in batch minority games with 2 strategies per agent;
(a) with completely uncorrelated strategies, (b) with each agent's 2 strategies 
mutually anti-correlated but with no correlation between agents. Shown are 
three different bias asymmetries between the points allocated initially to each 
agent's 2 strategies: $p_i(0)=0.0$ (circles), $0.5$ (squares) and $1.0$ (diamonds).
Also exhibited is a comparison betweeen the results of simulation 
of the deterministic many-agent dynamics (open symbols) and the 
numerical evaluation
of the analytically-derived stochastic single-agent ensemble dynamics. 
From \cite {GS03}.}
\label{figure1}
\end{figure}

Since the information on which the agents act is the same for all,
this problem is manifestly mean-field. It therefore offers the
potential for exact solution for its macro-behaviour in the sense of
the elimination of the microscopic variables in favour of
self-consistently determined macro-parameters in the 
limit of large $N$ \cite{SK}. The physics seems robust to variations of detail, but 
for completeness we indicate the version discussed explicitly.

Each agent $i,\, i=1,\dots,N$, is taken to
have two $D=dN$-dimensional strategies
$\bR_{ia}=(R_{ia}^1,\dots,R_{ia}^{dN}),\, a=\pm1$, with each component
$R_{ia}^\mu$ chosen independently randomly $\pm 1$ at the outset and
thereafter fixed. The common random information enters in that 
$\mu(t)$ is chosen stochastically randomly at each
time-step $t$ from the set $\mu(t)\in\{1,\dots,D\}$ and each agent
plays one of his/her two strategies $R_{ia_i}^{\mu(t)},\,a_i=\pm
1$. The actual choices of ${a_i}$ used, ${b_i(t)}$, are determined by the
current values of point differences ${p_i(t)}$. Let us restrict initially to
deterministic choices, $b_i(t)=\mbox{sgn}(p_i(t))$. The $p_i(t)$
are updated every $M$ time-steps according to
\begin{equation}\label{onlineupdate}
\hspace{-0.5cm}
p_i(t+M)=p_i(t)-M^{-1}\sum_{\ell=t}^{t+M-1}
\xi_i^{\mu(\ell)}\left\{N^{-1/2}\sum_j(\omega_j^{\mu(\ell)}+\xi_j^{\mu(\ell)}\mbox{sgn}(p_j(t)))\right\}
\end{equation}
where
$\bomega_i=(\bR_{i1}+\bR_{i2})/2,\,\bxi_i=(\bR_{i1}-\bR_{i2})/2$. 
In the so-called `online' game $M=1$ but here we consider the 'batch'
game where   $M\geq O(N)$ so
that the sum on the actual
$\mu(\ell)$ in (\ref{onlineupdate}) may be replaced by an average \cite{GS05} 
so that \cite{F2} \cite{F1} 
\begin{equation}\label{batchupdate}
\hspace{-0.5cm}
p_i(t+1)=p_i(t)  -\sum_{j} J_{ij} \mbox{sgn}{(p_j(t))}- h_i \equiv p_i(t) - \partial{H}/\partial{s_i}\mid_{s_i = \mbox{sgn}(p_i(t))};
\end{equation}
where  $J_{ij} = \sum_{\mu=1}^{D}{\xi_{i}^\mu}{\xi_{j}^\mu}$, $h_{i} = \sum_{\mu=1}^{D}
{\omega_{i}^\mu}{\xi_{i}^\mu}$ and 
$H = \sum_{(ij)} J_{ij}s_{i}S_{j} + h_{i} s_{i}$.

To proceed we use the dynamical generating functional method \cite{Coolen05}
with
\begin{equation}\label{generatingfunctional}
\hspace{-0.5cm}
Z =\int {\prod_{t}} d{\bf p}(t) W({\bf p}(t+1) \mid {\bf p}(t)) P_0 ({\bf p}(0)), 
\end{equation}
where ${\bf p}(t)=(p_1(t),\dots,p_N(t))$, 
$W({\bf p}(t+1)\mid {\bf p(t)})$ denotes the transformation 
operatation of eqn. \ref {batchupdate} and  $P_0({\bf p}(0))$ denotes the probability distribution of the 
initial score differences  from which the 
dynamics is started.
We consider the typical case by averaging over the specific choices of quenched strategies.
The averaged generating functional may then be transformed exactly into a form involving 
only macroscopic but temporally non-local variables (${\bf{\tilde{C}}}$, ${\bf{\tilde{G}}}$ and 
${\bf{\tilde{K}}}$) relatable to the 
correlation and response functions
of the original many-agent problem:
\begin{equation}\label{extremalfunctional}
\hspace{-0.5 cm}
Z=\int {D\tilde{C}(t,t') D\tilde{G}(t,t') D\tilde{K}(t,t') \exp\left(N \Phi({{\bf{\tilde{C}}}, 
{\bf{\tilde{G}}}, {\bf{\tilde{K}}}})\right)},
\end{equation}
where $\Phi$ is independent of $N$ and the bold-face notation denotes matrices in time. 
This expression is extremally dominated in the large $N$ limit and steepest descents yields
an effective stochastic single agent dynamics
\begin{equation}\label{effectiveagent}
\hspace{-0.5 cm}
p(t+1) = p(t) -\alpha \sum_{t' \leq t} ({\bf{1}} + {\bf{G}})^{-1}_{tt'} \mbox{sgn}p(t') + {\sqrt \alpha} \eta(t),
\end{equation}
\begin{equation}\label{colourednoise}
\hspace{-0.5 cm}
{\mbox{where}} \; \; \; \langle \eta(t) \eta(t') \rangle = [({\bf{1}} + 
{\bf{G}})^{-1}({\bf{1}}+{\bf{C}})({\bf{1}}+{\bf{G^T}})^{-1}]_{tt'}
\end{equation}
and the {\bf{G}} and {\bf{C}} are two-time response and correlation functions 
be determined self-consistently as averages over an 
ensemble of such single agents \cite{F4}; see \cite{Coolen05} for details. 
In the limit of large $N$ this analysis
is believed to be exact, but it is highly non-trivial. Empirical evidence is 
shown in fig 1 where comparison is made between the results of 
simulations over many 
instances of the  many-agent eqn. \ref{batchupdate} and numerical evaluations 
of the analytically-derived single-agent dynamics of 
eqn. \ref{effectiveagent}, including extension to anti-correlated strategies  \cite{GS05}. 

{\em Hence, naive characterization in terms
of a unique deterministic `representative agent', in the 
sense of conventional economics theory, is not possible. However, a 
single effective-agent description is available in 
a much more subtle sense. This is that one can consider the system to 
behave as though one has a {\bf{`representative stochastic ensemble'}} of 
non-interacting agents experiencing memory-weighting and coloured noise,
both determined self-consistently over the ensemble. Note 
that eqn. \ref{effectiveagent} is stochastic even though 
eqn. \ref{batchupdate} is deterministic.}

To go further one would need to solve the effective single-agent 
ensemble dynamics in a closed form. A complete soultion is 
not currently possible. However, one can solve for certain 
quantities in the ergodic equilibrating region. This concerns the aymptotic long-time behaviour, which is stationary
so that the two-time correlation and response functions become 
functions only of the relative times (ie. 
$G(t,t')$ and $C(t,t')$ become functions only of $(t'-t)$). Assuming also finite integrated response and 
weak long term memory leads to a formulation determining self-consistently the asymptotic order 
parameters $Q=\lim_{\tau \rightarrow \infty}C(\tau)$ and the integrated response $\chi=\sum_{\tau} G(\tau)$. Breakdown 
of the ergodic regime is signalled by diverging integrated response. Again the analytic theory works well 
within this ergodic regime, as is demonstrated in fig 2. 
The volatility, however, requires also the non-stationary parts of $C$ and $G$ and remains incompletely
solved in general, even in the ergodic regime \cite{GS05}.

\begin{figure}
\centerline{\includegraphics[width=30pc]{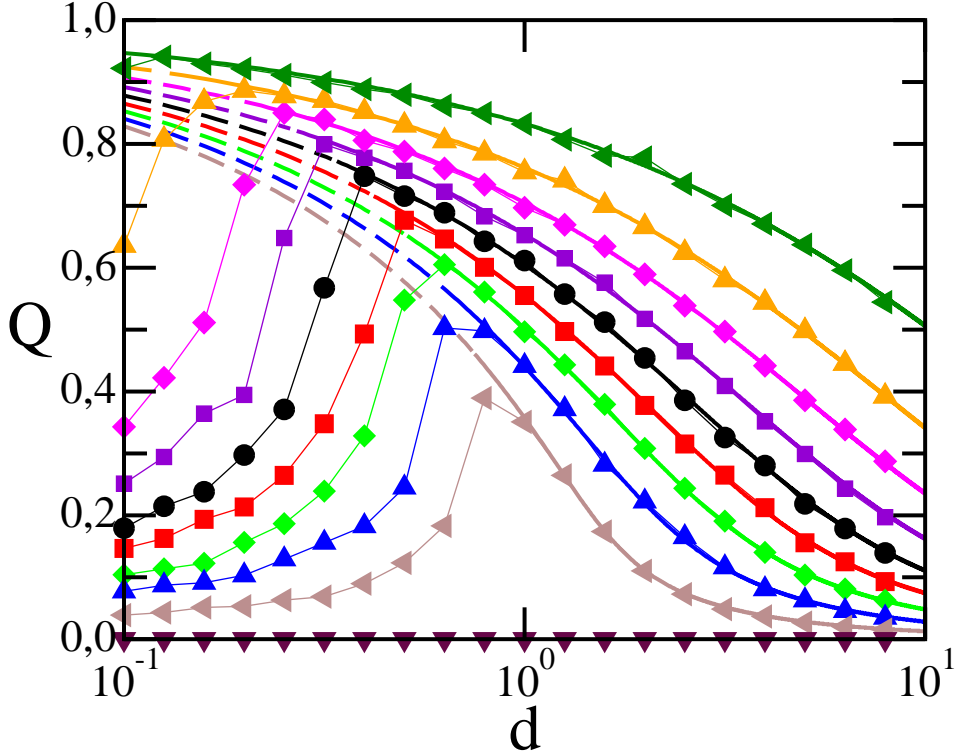}}
\vspace{1em}
\caption{Persistent part $Q$ of the correlation function for the
batch MG with {\em tabula rasa}
initial conditions. Symbols are simulation data. Solid lines are the
theoretical predictions for the ergodic regime, extrapolated
as dashed lines into the non-ergodic phase below $d_c$ (where
they are no longer valid), the changeover
signalling the predicted breakdown of the ergodic assumption. The 
different curves are
for different degrees of mutual correlation between agents' two
stategies; from anticorrelated at the bottom to highly correlated at the top. From\cite{GS05}.}
\label{figure2}
\end{figure}
 
The origin of the large volatilities found for {\em{tabula rasa}} starts in the non-ergodic 
regime can be ascribed to oscillatory behaviour, clearly visible empirically 
for such starts in quantities like the 
temporal correlation function
\begin{equation}
C(\tau)= \lim_{t\to\infty}
N^{-1}\sum_i\mbox{sgn}(p_i(t+\tau))\mbox{sgn}(p_i(t)),
\end{equation}
which exhibits persistent oscillations (with period 2 in the rescaled 
time units of eqn. \ref{batchupdate}) for $d<d_c$ \cite{GS05}. {\em Tabula rasa} starts 
in this region exhibit essentially no frozen agents, whereas highly biased starts 
result in mostly frozen agents and hence reduce the oscillations and with them 
the excess time-averaged volatility. The oscillations and the excess 
volatility are also reduced by random asynchronous point updating  \cite{GS05} and
by adding appropriate stochasticity to the original MG dynamics \cite{CGGS}\cite{CHS}.

Thus, as well as its possible relevance as an idealized economics model,
the Minority Game is of interest as a novel complex 
many-body system with both similarities and differences 
compared with other problems previously studied in statistical physics. 
Techniques developed within the spin glass community have proven useful in its analysis
and suggest extensions to other dynamical many-body systems characterised 
by a  combination of local/personal and global/range-free parameters, such as are typfied
by stockmarkets (and in contrast to those of most conventional condensed matter systems), 
without the need for Markovian or detailed balance dynamics. A complete solution to 
the effective single-agent stochastic ensemble remains still a challenge.

{\bf Acknowledgements}

Thanks are due to A. Cavagna, T. Coolen, T. Galla, 
J. Garrahan, I. Giardina, A. Heimel and E. Moro for collaborations and to EPSRC, ESF
 (SPHINX) and EC (STIPCO) for financial support.


\begin{thebibliography}{99}
\bibitem{F2} The time unit is rescaled here.
\bibitem{F1} Hence the system behaves as if controlled by a `Hamiltonian' 
of spins interacting through random exchanges and in random fields.
\end{thebibliography}
\end{document}